\documentstyle[psfig,epsf,conf-X,10pt]{article}
\begin{document} 
\small
\heading{%
%Begin Heading
%
ROSAC: Studying the clustering properties of X-ray selected AGNs
% End Heading
}
\par\medskip\noindent
\author{%
%Begin Author names
F. Tesch$^{1,2}$, F. J. Carrera$^{3,4}$, D. Engels$^{2}$, J. Hu$^{5}$, 
C. Ledoux$^{6}$, A. Ugryumov$^{7}$, D. Valls-Gabaud$^{8}$, 
W. Voges$^{9}$, J. Wei$^{5}$
%End Author names
}
\address{%
%First address
Harvard-Smithsonian Center for Astrophysics
}
\address{%
% Second Address
Hamburger Sternwarte
}
\address{%
% Third Address
Mullard Space Science Laboratory UCL
}
\address{%
% Fourth Address
Instituto de Fisica de Cantabria, CSIC-UC
}
\address{%
% Fifth Address
Beijing Astronomical Observatory
}
\address{%
% Sixth Address
Observatoire de Strasbourg
}
\address{%
% Seventh Address
Special Astrophysical Observatory RAS
}
\address{%
% Eigth Address
Observatoire Midi-Pyrenees, Toulouse
}
\address{%
% Nineth Address
MPE Garching
}

\begin{abstract}
We present the first ROSAC results of an AGN clustering analysis. 
This study comprises
a sample of 200 AGNs, 75\% of which being at low redshifts z$<$0.5, in the 
Ursa Major constellation.
The spatial 2-point-correlation function (SCF) as well as the minimal 
spanning tree (MST) technique were applied. 
Some evidence for clustering is found in the SCF, although with low 
significance.
Using the MST technique, we could find two AGN groups. This result is 
preliminary and the exact significance will be tested with careful
simulations.
\end{abstract}
\section{ROSAC: A ROSAT based Search for AGN Clusters}
The ROSAT All-Sky Survey (RASS) provides an excellent opportunity to study 
AGNs at low-redshifts. For the identification of RASS sources, objective 
prism and direct plates from the Hamburg Quasar Survey 
were used, giving a list of AGN candidates. The AGN 
nature of these candidates has to be confirmed by follow-up spectroscopy. 
Our confirmation rate for AGN candidates is $\approx$ 95\%, which makes
this identification strategy powerful for creating AGN samples. 

The ROSAC project makes use of this work to study the spatial properties
of low-redshift AGNs. In particular the search for clusters or groups of AGNs
and the determination of the 2-point-correlation function come to the fore.
Three regions in the
constellations Ursa Major (UMa), Coma Berenices, and Pisces were selected
due to a) low hydrogen column densities, b) large numbers of known
redshifts to reduce the observing time, and c) the presence of interesting
structures found in a first minimal spanning tree analyses. 
The most advanced 'subsample' today is that in UMa with a 
completeness 
of 87\%. This region covers an area of 363 deg$^{2}$ and consists of 200 
confirmed AGNs. A first clustering investigation within the scope of the 
ROSAC project is restricted to UMa (Fig. 1).

\section{Clustering Analyses}
The 2-point-correlation function $\xi(r)=\frac{N_{AGN}}{N_{Random}}-1$ was
applied. Clustering properties of AGNs in the low redshift regime are 
uncertain. 
Only a few investigations of small samples and with lower surface densities 
than the ones of the ROSAC project have been conducted so far. 
Specifically, the two studies of X-ray selected AGNs (Boyle \& Mo 1993, 
Carrera et al. 1998) did not show clustering on small scales. 

The results of our investigation are outlined in Fig 1. A power law 
$\xi(r)=(\frac{r}{r_{0}})^{-\gamma}$ was fitted to the data. The 
correlation length gives $r_{0}=8.1^{+2.7}_{-3.9}$ and 
$\gamma=1.08^{+0.45}_{-0.23}$, which is consistent with the favoured value 
of $r_{0}=6.0$ for AGNs.

Additionally, a search for groups of AGNs was carried out using the 
minimal spanning tree (MST) technique. Earlier studies could find 18 groups of 
AGNs in total. All of these groups were confirmed by the MST technique.
We could detect two further AGN groups comprising 21 and 14 members at mean
redshifts of 0.081 and 0.222. Further work is necessary to quantify the exact
significance of these results.
\begin{figure}
%\mbox{\epsfxsize=10cm \epsffile{fig.ps}}
\begin{minipage}[t]{11cm}
 \begin{minipage}[t]{5cm}
  \begin{flushleft}
   \psfig{figure=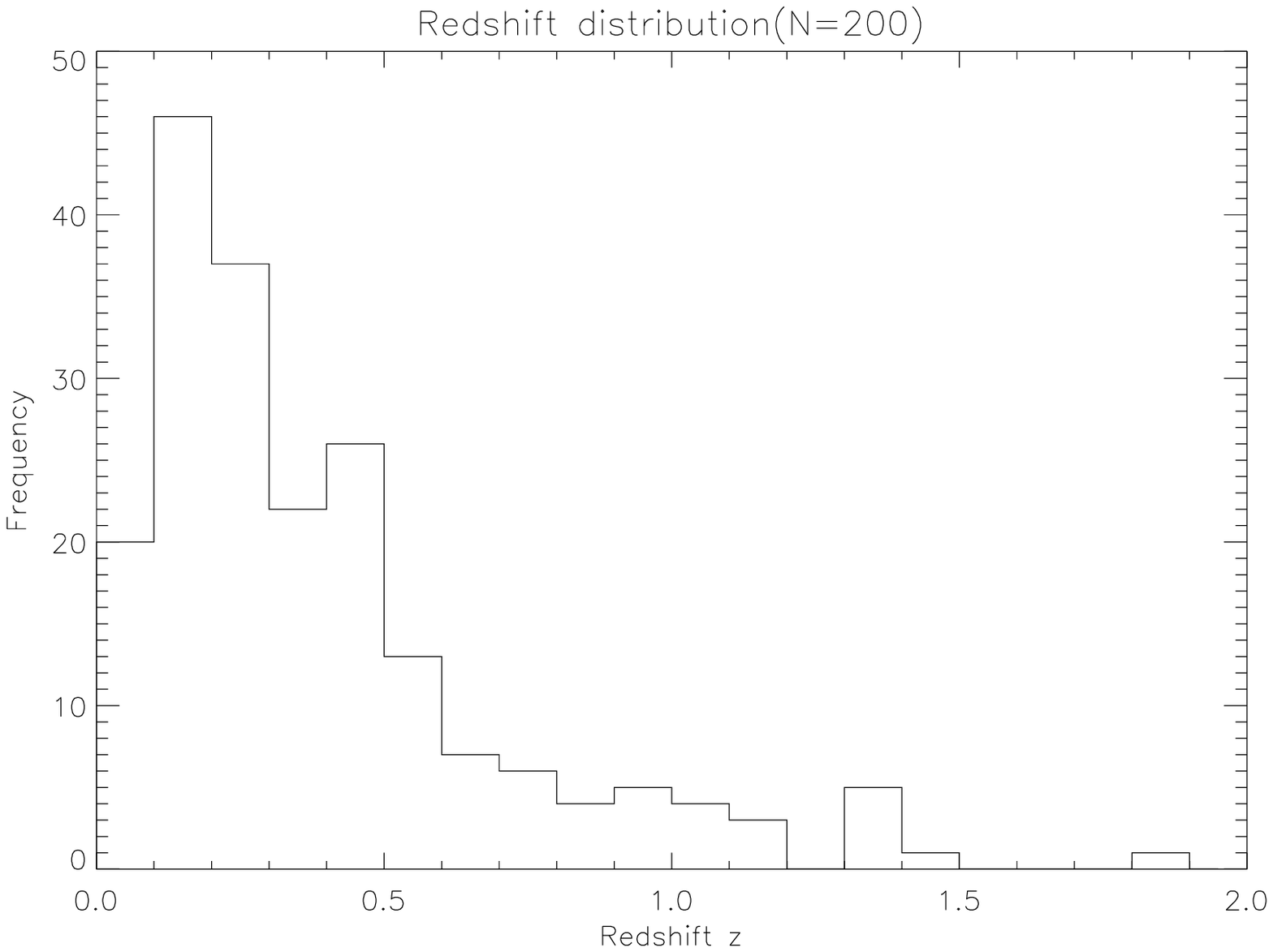,width=5cm}
  \end{flushleft}
 \end{minipage}
 \hfill
 \begin{minipage}[t]{5cm}
  \begin{flushright}
   \psfig{figure=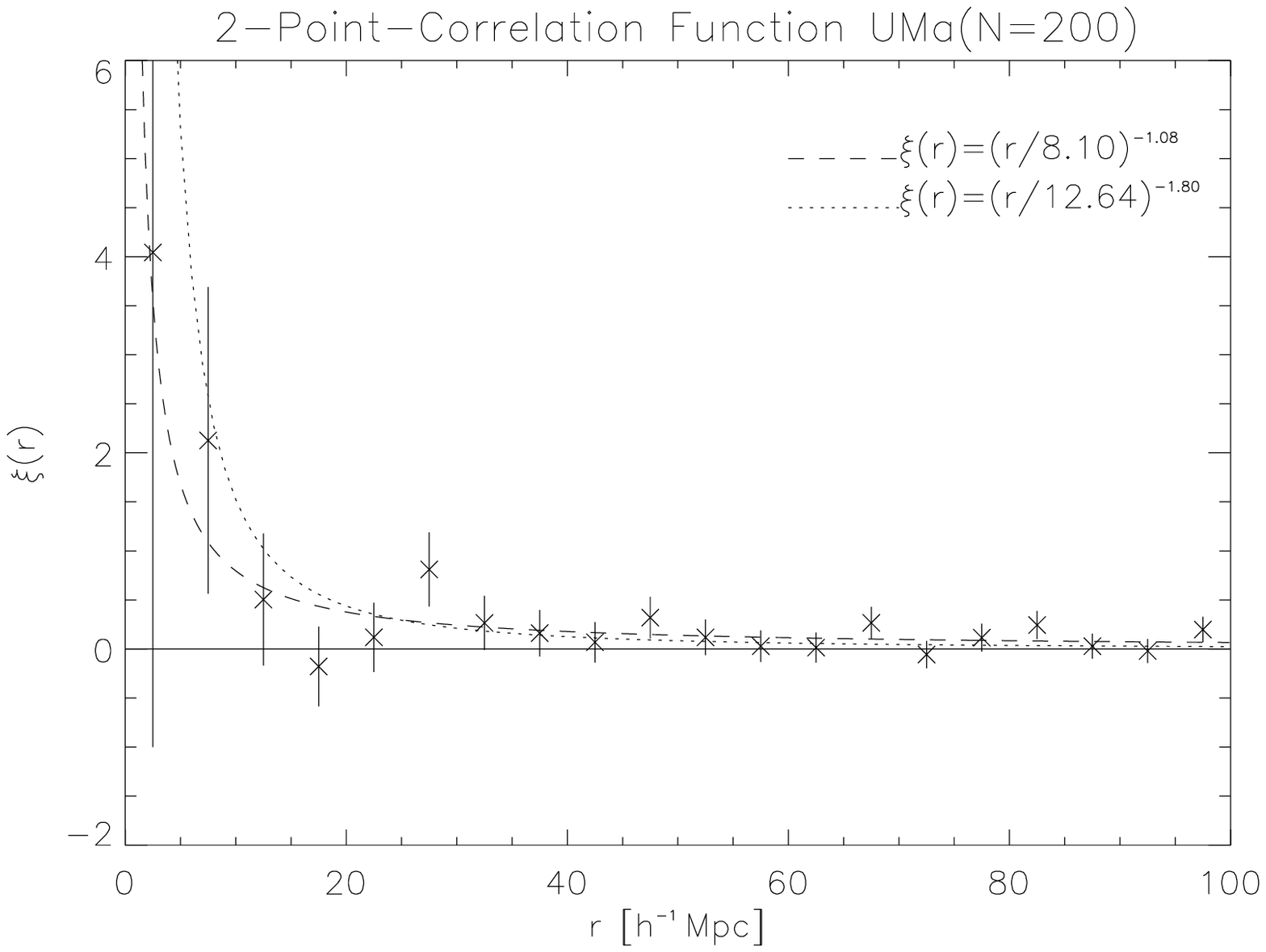,width=5cm}
  \end{flushright}
 \end{minipage}
\end{minipage}
%
%\begin{figure}
%\centerline{\vbox{
%\psfig{figure=fig1.ps,height=7.cm}
%}}
\caption[]{
Redshift distribution and 2-point-correlation function 
$\xi(r)=(\frac{r}{r_{0}})^{-\gamma}$: 
$\chi^{2}$-fits with $r_{0}$ as free parameter (dotted line) and with two
free parameters (dashed line) are shown. The error bars are poissonian.
}
\end{figure}
\section{Future prospects}
The final ROSAC project will result in a sample of about 700 AGNs with surface 
densities between 0.3 and 0.5 AGNs/deg$^{2}$ at redshifts z$<$0.5.
Consequently, the incorporation of the two other regions, which means 
500 objects additionally, would provide a much better sample to study AGN
clustering.
%
%

%
%\section{References}
%\begin{iapbib}{99}{
%\bibitem{Kea} Keaton B., 1927, \aeta 555, 556
%\bibitem{LH} Laurel S., Hardy O., 1931, \apj 38, 357
%\bibitem{MMM} Marx G., Marx H., \& Marx C., 1938, eds Chaplin C.,
%              in {\it Our lives}. MGM Editions, Paris, p. 129
%}
%\end{iapbib}
%
%\vfill
\end{document}